\documentclass[12pt]{article}
\usepackage{amsmath}
\usepackage{amssymb}
\usepackage{graphicx,psfrag,epsf}
\usepackage{enumerate}
\usepackage{natbib}
\usepackage{subcaption}
\usepackage[ruled]{algorithm2e}
\usepackage{url} 

\newcommand{\blind}{1}

\begin{document}

\def\spacingset#1{\renewcommand{\baselinestretch}%
{#1}\small\normalsize} \spacingset{1}


\if1\blind
{
  \title{\bf A Non-homogeneous Count Process: Marginalizing a Poisson Driven Cox Process}
  \author{Shuying Wang\thanks{
  For correspondence:
    shuying.wang@utexas.edu}\hspace{.2cm}\\
    Department of Statistics and Data Sciences, University of Texas at Austin\\
    and \\
    Stephen G. Walker\thanks{s.g.walker@math.utexas.edu}\\
    Department of Mathematics, University of Texas at Austin}
    \date{}

    \maketitle
} \fi

\if0\blind
{
  \bigskip
  \bigskip
  \bigskip
  \begin{center}
    {\LARGE\bf Title}
\end{center}
  \medskip
} \fi

\abstract{The paper considers a Cox process where the stochastic intensity function for the Poisson data model is itself a non-homogeneous Poisson process. We show that it is possible to obtain the marginal data process, namely a non-homogeneous count process exhibiting over-dispersion.
While the intensity function is non-decreasing, it is straightforward to transform the data so that a non-decreasing intensity function is appropriate. We focus on a time series for arrival times of a process and, in particular, we are able to find an exact form for the marginal probability for the observed data, so allowing for an easy to implement estimation algorithm via direct calculations of the likelihood function.}

\vspace{0.2in}

\spacingset{1.2}
\section{Introduction}

The family of Cox processes are non--homogeneous Poisson processes  with a random intensity function. Differences arise in how this random function is constructed. Original applications were the modeling of heterogeneous spatial and/or spatio--temporal point processes, such as \cite{wolpert1998}. Recent developments relied on the log--Gaussian Cox process; see \cite{brix2001, diggle2013, teng2017, bayisa2020}. The rationale behind the doubly stochastic process is to deal with the issue of over-dispersion. As is well known, the mean and variance of a Poisson random variable coincide, contrary to many observed phenomenon in which the variance of a point process exceeds the mean.

In a spatial setting, and even in a one dimensional time sequence case, there is a need to employ a covariance structure to the modeling of the intensity function. Hence, the Gaussian process is the main tool, though, due to the need for a positive intensity function, it is a log--Gaussian process which is used. In fact, any transform of a Gaussian process to a positive function would be possible. Other types of model for the intensity include a shot noise process and other Markov point processes. The original work on random intensity functions include \cite{thomas1949} and \cite{neyman1958}, who used simplified shot--noise type processes. 

While the applications for the Cox process are predominantly to do with spatial modeling, we focus attention in this paper to time series data, so we will only be considering random intensity functions which exist on $\left(0,\infty\right)$. This was also the setting for the work of \cite{Wu2013}, though these authors assumed they observed repeated point processes, and used functional data analysis techniques, whereas we assume only a single process is observed.

To set the model and notation, and we will focus on the one dimensional setting, indexed by time $t$, 
$X(\cdot)$ conditional on the random  process $Y(\cdot)$  is a Poisson process with intensity function $Y(t)$.
The requirement  for the existence of the Cox process is that
$\int_0^T Y(s)\, ds<\infty$
for all $T<\infty$.  

A popular and original form for the intensity process is given by
$$Y(t)=\sum_{i}w_i\,f(t;s_i),$$
for some set of time points $S=(s_i)$, and $f$ is a probability density function with $w_i$ the weight for point $s_i$.
This is the general form for the shot noise Cox process. 
The mean intensity function is given by
$$\rho(t)=\int\int w\,f\left(t;s\right)\,ds\,\chi(dw)$$
for some density $\chi$ on $(0,\infty)$. 
The \cite{neyman1958} class of process has $w_i=w$ and the \cite{thomas1949} process when $f$ is based on the Gaussian density. 

The process is stationary when $f(t;s)=f(t-s)$and in this case the mean function is 
$\rho(t)=\rho=\int w\,\chi(w)\,dw.$ 
For more on the shot noise Cox process, see \cite{moller2003, moller2005, dassios2015, jalilian2015} and for recent developments including the determinantal shot Cox process, see \cite{moller2022}, while for a comprehensive review of the Cox process and associated processes, see \cite{Jang2021}.

On the other hand, the log Gaussian process, see \cite{moller1998}, arises by taking
$Y(t)=\exp\{Z(t)\}$
with $Z$ a Gaussian process; that is for any set of points $(t_1,\ldots,t_m)$, it is that $(Z(t_1),\ldots, Z(t_m))$ has a multivariate  Gaussian distribution. 
There are by now many papers devoted to the log Gaussian Cox process, both theoretical and algorithms for estimating the model. Recent advances for the latter necessarily involve Markov chain Monte Carlo (MCMC) techniques. A grid becomes necessary and hence a discretized approximation of the Gaussian process on a set of points is used; see for example, \cite{Shirota2016}. 
Other types of process for $Y$ appear in \cite{schnoerr2016}.

In this paper we model $Y(t)$ in a novel way, and which allows us to find the marginal non-homogeneous count process for $X$. We have $Y$ to be non--decreasing and indicate how we can model arbitrary count data using such a random intensity function. 
We take $Y(t)$ to be driven by a non--homogeneous Poisson process; that is
$$Y(t)=\sum_{i}1(s_i\leq t),$$
where the $(s_i)$ are from a non--homogeneous Poisson process with intensity function $\gamma(t)$. Hence, the mean intensity function for $X(t)$ is
$\rho(t)=\int_0^t\gamma(u)\,du.$
The important result from the work in this paper is that it is possible to derive the marginal joint probability for the $X$ process. That is, if $X$ on $[0,T]$ has $m$ changes at time locations $t_1<\cdots <t_m$, then we are able to provide an expression for 
$p(t_1,\ldots,t_m,m).$
This ensures an easy to implement inferential algorithm which needs no approximation based on a discretization and no sampling of a latent stochastic process.


Given the $Y$ process lives on particular states, i.e. the non-negative integers, the model can also be regarded as a continuous time hidden Markov model where the hidden Markov part of the model is non-homogeneous. These models are notoriously difficult to estimate, even when the hidden continuous time process is homogeneous and lives on a finite number of states. The usual recourse is to discretize time and/or to use an EM algorithm for maximizing the likelihood. Even these ideas are difficult to implement. Nevertheless, for a hidden non-homogeneous Poisson process we are able to provide an expression for the likelihood function which can be easily computed. While we could regard the model as a non-homogeneous continuous time hidden Markov model, we refer to it as coming from the family of Cox processes.

The rest of the paper is organized as follows. In Section~\ref{model} we present the form of the model in terms of infinitesimal probabilities, define the marginal probability of the $X$ process by the expectation $\text{E}_Y\left[p_{X|Y}(x\mid Y)\right]$,
and also lay down the notation to be used throughout the paper. In Section~\ref{discretization} 
we describe a discrete-time count path $x^{(n)}$ and a discrete-time count process $Y^{(n)}$ as the $n$-grid discretization of $x$ and $Y$. 
In Section~\ref{derivation} we derive the expectation $\text{E}_{Y^{(n)}}\left[p_{X|Y}\left(x^{(n)}\mid Y^{(n)}\right)\right]$ by recursion. In Section~\ref{convergence} we prove that $\text{E}_{Y^{(n)}}\left[p_{X|Y}\left(x^{(n)}\mid Y^{(n)}\right)\right]$ converges to $\text{E}_Y\left[p_{X|Y}(x\mid Y)\right]$ as $n$ goes to infinity and thus take limit of this expectation to get an analytical expression of the marginal probability of $X$ and hence the likelihood.

\section{Model and Background}\par
\label{model}
Consider a Cox process $\{X(t)\}_{0\le t\le T}$, which is a non-homogeneous Poisson process with a random intensity function $\beta(t,\  Y(t)) = b(t) + wY(t)$, where $\{Y(t)\}_{0\le t\le T}$ is a non-homogeneous Poisson process with intensity function $\gamma(t)$. In this paper, we take $b(t) = \beta_0$ to be a non-negative constant, but it is possible to be generalized to any real integrable function. 

The infinitesimal probabilities of $X$ conditional on $Y$ are given by
\begin{equation*}
\begin{cases}
\text{P}\left(X(t+h)=x+1\mid X(t)=x,\ Y(t)=y\right)=\beta(t,\  y) h+o(h) \\
\text{P}\left(X(t+h)=x\mid X(t)=x,\ Y(t)=y\right)=1 - \beta(t,\  y) h+o(h), 
\end{cases}
\end{equation*}
and the infinitesimal probabilities of $Y$ are given by
\begin{equation*}
\begin{cases}
\text{P}\left(Y(t+h)=y+1\mid Y(t)=y\right)=\gamma(t)h+o(h) \\
\text{P}\left(Y(t+h)=y\mid Y(t)=y\right)=1 - \gamma(t) h+o(h). 
\end{cases}
\end{equation*}
Our aim is to find the marginal process $\{X(t)\}_{0<t\le T}$ by integrating out the $Y$ process. To see that this is a non-trivial exercise, consider the simplest version of the model; i.e.
$X(t)$ given $Y$ is Poisson with mean $\int^t_0 Y(s)\, ds$ and $Y(t)$ is a Poisson process with mean $t$. The Laplace transform of $X(t)$ is given by
$$-\log E e^{-\theta X(t)}=\phi^{-1}(1-e^{-\phi t}),$$
where $\phi=1-e^{-\theta}$. This is not recognizable for the distribution of $X(t)$, let alone for the process.

The right continuous sample paths of $\{X(t)\}_{0\le t\le T}$ and $\{Y(t)\}_{0\le t\le T}$, denoted by $x:\left[0,T\right]\rightarrow \mathbb{N}$ and $y:[0,T]\rightarrow \mathbb{N}$, can be characterized by the time points of the jumps. Let $\{t_i\}$ be the time points of jumps in $x$, and $\{s_i\}$ be the time points of the jumps in $y$, then
$$x(t) = \sum_{t_i} \mathbf{1}(t_i\le t),\ \ \ y(t) = \sum_{s_i} \mathbf{1}(s_i\le t),\ \ \ \forall\ t\in[0, T].$$
The conditional distribution of $X$ given $Y$ is given by
\begin{equation}
\label{pxgiveny}
p_{X|Y}(x\mid y) = \prod_{t_i}\ \beta(t_i-,\ y(t_i-))\ \text{exp}\left\{-\int_0^T\beta(t,\ y(t))\ dt\right\},
\end{equation}
and the marginal distribution of $Y$ is given by
$$p_Y(y) = \prod_{s_i}\ \gamma(s_i-)\ \text{exp}\left\{-\int_0^T\gamma(t)\ dt\right\}.$$
The marginal likelihood of $X$ can be expressed as an expectation with respect to $Y$,
$$p_X(x)=\int p_Y(y)\ p_{X|Y}(x\mid y)\ dy=\text{E}_Y\left[p_{X|Y}(x\mid Y)\right],$$
which is the expected value of (\ref{pxgiveny}) with respect to $y$. 
It is hard to find the expectation directly, since $Y$ is a continuous-time process. The only simple case is when $m=0$ for which
$$p_X(x)=\exp\left\{-\int_0^T(1-e^{-(T-s)})\gamma(s)\,ds\right\},$$
where we have also taken $w=1$ and $\beta_0=0$ to get this expression.

Our approach is to perform the integration on an $n$-grid discretization for both $x$ and $Y$, and derive the expectation $\text{E}_{Y^{(n)}}\left[p_{X|Y}\left(x^{(n)}\mid Y^{(n)}\right)\right]$ with a discrete-time count path $x^{(n)}$ and discrete-time count process $Y^{(n)}$. We then take the limit of the expectation as $n\rightarrow\infty$ to recover the desired expectation of the limit. The exchange of the expectations and the limit, and hence the convergence of expectations, can be shown by the convergence of $x^{(n)}\rightarrow x$ in the Skorokhod topology and the convergence of $Y^{(n)}\rightarrow Y$ in distribution. 

The plan for finding the marginal probability of a number of events $m$ with corresponding event times is as follows. We find the answer for $m=0$ for the discretized process. 
We then use an induction argument to find the likelihood function for general $m$ and, once done, we allow the discretized process to be returned to a continuous process and find the corresponding limits for the marginal probability. 

To explain more clearly the discretization procedure, suppose $Y_n$ is a sequence of random variables such that $Y_n$ converges in distribution to random variable $Y$. So $Y_n$ could be a discretization of $Y$ (of outcomes) or some other approximation to it.
Then, as is well known, for all $g$ continuous and bounded, it is that
$$\lim_{n\to\infty} E(g(Y_n))=E(g(Y)).$$
In short, we are estimating $E(g(Y))$ using the limit of the $E(g(Y_n))$, and applying this idea when the $Y$s are certain stochastic processes. Hence, we start with a description of the discretization (in time)/approximation of the processes involved. 
Then we find the expectation of the $n$ approximation and finally obtain the limit as $n\to\infty$.

\section{Discretization of the Processes}\par
\label{discretization}
The expectation $\text{E}_{Y^{(n)}}\left[p_{X|Y}\left(x^{(n)}\mid Y^{(n)}\right)\right]$
is taken with respect to a discrete-time point process $Y^{(n)}$ and with a discrete-time count path $x^{(n)}$, but the conditional probability function $p_{X|Y}(\cdot\mid\cdot)$ has not been discretized. Therefore, in this section we will define $x^{(n)}$ and $Y^{(n)}$ as the $n$-grid discretization of the count path $x$ and non-homogeneous Poisson process $Y$. 

Let $D[0, T]$ be the space of real functions on $[0, T]$ that are right-continuous with left-hand limits. Furthermore, let $D_c[0, T]$ be the set of count paths in $D[0, T]$. 
A function in $D$ is called a count path if ``it is non-decreasing, take integers as values, and has jumps of exactly 1 at its points of discontinuity" \citep{billingsley2}, so count paths are the sample paths of point processes.

Take the sample path $x$ as defined in Section~\ref{model}, so $x\in D_c[0, T]$. For each $x$, denote its $n$-grid discretization as $x^{(n)}$. Let
$x^{(n)}$ agree with $x$ at the discrete time points in $\mathcal T^{(n)} := \left\{k\,T/n:k=1,\ldots,n\right\}$ and stays constant in each time interval $[(k-1)T/n, kT/n)$ for $k = 1,\ldots,n$, so $x^{(n)}$ is also a right continuous count path in $D_c[0, T]$, but can only jump at $t\in \mathcal T^{(n)}$. Here we don't consider the case when $x^{(n)}$ has two jumps at the same time, because it won't be a problem for large $n$, as long as the $n$-grid is fine enough. 
Denote the time points of jumps in $x^{(n)}$ as $\left\{t_i^{(n)}\right\}$, so $t_i^{(n)} = \min\left\{s\in \mathcal T^{(n)}:s\ge t_i\right\}$.

Take the non-homogeneous Poisson process $Y$ as defined in Section~\ref{model} and denote its $n$-grid discretization as $Y^{(n)}$.
Let $Y^{(n)}$ be a point process that can only jump at the discrete time points $t \in \mathcal T^{(n)}$, with probability $\gamma(t)\,T/n$, so we have
\begin{align*}
& \text{P}\left(Y^{(n)}\left(t+T/n\right) = Y^{(n)}(t)+1\right) = \gamma(t)\,T/n & \forall\ t\in \mathcal T^{(n)},\ t\ne T.
\end{align*}
Now let $y^{(n)}$ be the sample path of $Y^{(n)}$. For notation simplification,  let $h=T/n$ be the size of the time grids. Let 
$x_k=x^{(n)}\left(kh\right)$ and $y_k=y^{(n)}\left(kh\right)$ for $k=0,\ldots, n$, so the sample paths $x^{(n)}$ and $y^{(n)}$ can be represented by $(x_0,\ldots x_n)$ and $(y_0,\ldots y_n)$.
We also simplify the notations of the intensity functions by
$\gamma_k = \gamma(kh)$ and $\beta_k =\beta(kh,\ y_k)$ for $k=0,\ldots, n$.

The probability function of $Y^{(n)}$ is given by
$$\text{P}_{Y^{(n)}}\left(y^{(n)}\right) = \prod_{k=0}^{n-1} \text{P}\left(y_{k+1}\ |\ y_k\right),$$
where 
\begin{align*}
\text{P}\left(y_{k+1}\ |\ y_k\right) =
\begin{cases}
     \gamma_kh & \text{for}\ \ y_{k+1} = y_k+1 \\
     1-\gamma_kh & \text{for}\ \ y_{k+1} = y_k.
\end{cases}
\end{align*}
When substituting the $x$ and $y$ in Equation~(\ref{pxgiveny}) by $x^{(n)}$ and $y^{(n)}$, the function $p_{X|Y}\left(x^{(n)}\mid y^{(n)}\right)$ can be written in a product form given by
\begin{align*}
    p_{X|Y}\left(x^{(n)}\mid y^{(n)}\right) & = \prod_{t_i^{(n)}}\ \beta\left(t_i^{(n)}-h,\ y^{(n)}\left(t_i^{(n)}-h\right)\right)\ \exp\left(-\sum_{k=0}^{n-1}\beta_kh\right) \\
    & =\prod_{k=0}^{n-1} p(x_{k+1}\ |\ x_k,\ y_k),
\end{align*}
where
\begin{align*}
    p(x_{k+1}\ |\ x_k,\ y_k) = 
    \begin{cases}
    \beta_k\exp(-\beta_kh) 
    &\text{for}\ \ x_{k+1} = x_k+1 \\
    \exp(-\beta_kh)  &\text{for}\ \ x_{k+1} = x_k.
    \end{cases}
\end{align*}
Let $\mathcal{Y}^{(n)}$ denote the set of all possible sample paths of $Y^{(n)}$. $\mathcal{Y}^{(n)}$ is countable even when $n\rightarrow \infty$ since $Y^{(n)}$ can only have finitely many jumps. Therefore, we can derive the expectation $\text{E}_{Y^{(n)}}\left[p_{X|Y}\left(x^{(n)}\mid Y^{(n)}\right)\right]$ by summing over all $y^{(n)}\in \mathcal{Y}^{(n)}$, when assuming $y_0=0$ is known,   
\begin{align*}
& \text{E}_{Y^{(n)}}\left[p_{X|Y}\left(x^{(n)}\mid Y^{(n)}\right)\right]  =  \sum_{y^{(n)} \in \mathcal{Y}^{(n)}} p_{X|Y}\left(x^{(n)}\mid y^{(n)}\right) \text{P}_{Y^{(n)}}\left(y^{(n)}\right) \\
 = & \sum_{y^{(n)} \in \mathcal{Y}^{(n)}}\ \prod_{k=0}^{n-1} p(x_{k+1}\mid x_k,\ y_k)\ \prod_{k=0}^{n-1} \text{P}(y_{k+1}\mid y_k) \\
 = &\  p(x_1\mid x_0,\ y_0) \ \sum_{y_1} p(x_2\mid x_1,\ y_1)\ \text{P}(y_1\mid y_0)  \ \sum_{y_2} p(x_3\mid x_2,\ y_2)\ \text{P}(y_2\mid y_1) \ldots\ldots \\
& \sum_{y_{n-1}} p(x_n\mid x_{n-1},\ y_{n-1})\ \text{P}(y_{n-1}\mid y_{n-2})\  \sum_{y_n} \text{P}(y_n \mid y_{n-1}).
\end{align*}\par
\ \par
To further simplify the notations, let's define the summation step by step,
\begin{align*}
& s_1(y_{n-1}) = \sum_{y_n} \text{P}(y_n \mid y_{n-1}) \\
& s_2(y_{n-2}) = \sum_{y_{n-1}}p(x_{n}\mid x_{n-1},\ y_{n-1})\ \text{P}(y_{n-1}\mid y_{n-2})\ s_1(y_{n-1})\\
& \ldots\ldots \\
& s_{k}(y_{n-k}) = \sum_{y_{n-k+1}} p(x_{n-k+2}\mid x_{n-k+1},\ y_{n-k+1})\ \text{P}(y_{n-k+1}\mid y_{n-k}) \ s_{k-1}(y_{n-k+1}) \\
& \ldots\ldots \\
& s_{n}(y_{0}) = \sum_{y_{1}}p(x_2\mid x_1,\ y_1)\ \text{P}(y_1\mid y_0)\ s_{n-1}(y_{1}). 
\end{align*}
Here the summation over $y^{(n)}\in\mathcal{Y}^{(n)}$ is going backwards along the timeline, so we first sum over $y_n$ to get $s_1(y_{n-1})$, and then sum over $y_{n-1}$ to get $s_2(y_{n-2})$, etc. When we derive the summation up to $s_k(y_{n-k})$, we have already summed over $y_{n}, y_{n-1},\ldots,y_{n-k+1}$, and have not yet reached $y_{n-k-1}$, so $s_k(y_{n-k})$ only depends on $y_{n-k}$. It also depends on $x_n,x_{n-1},\ldots,x_{n-k+1}$, so $s_k(y_{n-k})$ should actually be written as $s_k\left(y_{n-k},\,x^{(n)}\right)$, but here we put each $s_k(y_{n-k})$'s dependence on $x^{(n)}$ in silence, and only specify its dependence on $y^{(n)}$. We can try to find the inductive pattern between $s_k(y_{n-k})$ and $s_{k+1}(y_{n-k-1})$, to derive the summation to $s_n(y_0)$, and thus get the expectation given by
$$\text{E}_{Y^{(n)}}\left[p_{X|Y}\left(x^{(n)}\mid Y^{(n)}\right)\right] = p(x_1\mid x_0,\ y_0)\ s_{n}(y_0).$$
In the next section we find the marginal probability for the discretized process.

\section{Marginal Probabilities}
\label{derivation}

Here we derive $\text{E}_{Y^{(n)}}\left[p_{X|Y}\left(x^{(n)}\mid Y^{(n)}\right)\right]$ for a simple case, assuming $x_0=y_0=0$, and $\beta(t,\ y)=\beta_0+w y$, with $\beta_0\ge0$ and $w>0$, so $\beta_k = \beta(kh,\ y_k)=\beta_0 + wy_k$. We use an inductive argument on the number of jumps in $x^{(n)}$.

When deriving the summation at each step, we need to sum over $y^{(n)}$ and
consider whether or not there is a jump in $x^{(n)}$, at the corresponding time point. For example, the summation step given by
$$s_{n-k}(y_k)=\sum_{y_{k+1}}p(x_{k+2}\mid x_{k+1},\ y_{k+1})\,\text{P}(y_{k+1}\mid y_{k})\,s_{n-k-1}(y_{k+1})$$
is summing over $y_{k+1}$ and we need to consider $x_{k+2}$, so when there is no jump at $x_{k+2}$, i.e. $x_{k+2}=x_{k+1}$, we have
\begin{align*}
p(x_{k+2}\mid x_{k+1},\ y_{k+1})\,\text{P}(y_{k+1}\mid y_{k})=
    \begin{cases}
    \exp(-\beta_kh)\,(1-\gamma_kh) 
    &\text{for}\ y_{k+1} = y_{k} \\
    \exp\{-(\beta_k+w)h\}\,\gamma_kh  &\text{for}\ y_{k+1} = y_{k} + 1,
    \end{cases}
\end{align*}
and when there is a jump at $x_{k+2}$, i.e. $x_{k+2}=x_{k+1}+1$, we have
\begin{align*}
p(x_{k+2}|x_{k+1},\,y_{k+1})\text{P}(y_{k+1}|y_{k})=
    \begin{cases}
    \beta_k\exp(-\beta_kh)(1-\gamma_kh) 
    &\text{for}\, y_{k+1} = y_{k} \\
    (\beta_k+w)\exp\{-(\beta_k+w)h\}\gamma_kh  &\text{for}\, y_{k+1} = y_{k} + 1.
    \end{cases}
\end{align*}

\subsection{No Jumps}
\label{nojump}
First, assume $x^{(n)}$ is a sample path that has no jumps in the time interval $[0, T]$, so $x_{k} = x_{k-1}$ for all $k = 1,\ldots, n$.

The derivation of the summation starts from $s_1(y_{n-1})=\sum_{y_n} \text{P}(y_n \mid y_{n-1})=1$. Then we sum over $y_{n-1}$ to get
\begin{align*}
    s_2(y_{n-2})&=\sum_{y_{n-1}}p(x_{n}\mid x_{n-1},\ y_{n-1})\ \text{P}(y_{n-1}\mid y_{n-2})\\
    &=\exp(-\beta_{n-2}h)(1-\gamma_{n-2}h) + \exp\{-(\beta_{n-2}+w)h\}\gamma_{n-2}h\\
    &=\exp(-\beta_{n-2}h) \ \left\{1 -  \left( 1 - e^{-w h}\right)\gamma_{n-2} h\right\}.
\end{align*}
Next, we sum over $y_{n-2}$ to get
\begin{align*}
s_3(y_{n-3})& =\sum_{y_{n-2}}p(x_{n-1}\mid x_{n-2},\ y_{n-2})\ \text{P}(y_{n-2}\mid y_{n-3})\ s_2(y_{n-2})\\
&=\Big[\exp(-2\beta_{n-3}h)(1-\gamma_{n-3}h) + \exp\{-2(\beta_{n-3}+w)h\}\gamma_{n-3}h\Big]\ (1-\lambda_{n-2}h)\\
&=\exp(-2\beta_{n-3}h)\ ( 1-  \lambda_{n-2}h)\ (1 -   \lambda_{n-3}h),
\end{align*}
where 
$\lambda_{i}:=\left(1 - e^{-(n-i-1)w h}\right) \gamma_{i}.$
Keep repeating this, we can get, for $2\le k\le n$,
\begin{equation}
\label{0jump}
s_k(y_{n-k})=\exp\{-(k-1)\beta_{n-k}h\}\prod_{i=n-k}^{n-2} (1-\lambda_i h).
\end{equation}
The above equation can be easily proved by induction. We know that it's true for $s_2(y_{n-2})$, and it's also trivial to show that when it holds for $s_k(y_{n-k})$, it will also hold for $s_{k+1}(y_{n-k-1})$.

Therefore, at the end of the summation, we will get
$$\text{E}_{Y^{(n)}}\left[p_{X|Y}\left(x^{(n)}\mid Y^{(n)}\right)\right]=p(x_1\mid x_0,\ y_0)\ s_{n}(y_0)=\exp(-n\beta_0h)\ \prod_{i=0}^{n-2} (1-\lambda_i h).$$
With this as the starting point, we now find the corresponding form for $m$ jumps using an induction argument.

\subsection{Inductive Argument}
\label{mjumps}
In this section, we will derive a general expression of $\text{E}_{Y^{(n)}}[p_{X|Y}\left(x^{(n)}\mid Y^{(n)}\right)]$, by induction and recursion, for the sample path $x^{(n)}$ with any number of jumps in the time interval $[0, T]$.

Let us denote the jump times as $t_i^{(n)}=(r_i+2)h$, labeled in a descending order, with $t_i^{(n)}>t_j^{(n)}$ for $i<j$. We first derive a general expression of $s_k(y_{n-k})$ by induction.\par
\medskip
\noindent
\textbf{Theorem 1}.
Given a sample path $x^{(n)}$, let $m$ be the number of jumps in $x^{(n)}$ that has been counted in the summation up to $s_k(y_{n-k})$, i.e. $m = \sum_{r_i}\mathbf1(r_i \ge n-k)$. Then, there exist a set of coefficients $\left\{c_{j,\ k}^{(m,\,n)}\right\}_{j=0}^m$ that only depends on $x^{(n)}$ but does not depend on $y^{(n)}$, such that
the general expression of $s_k(y_{n-k})$, for $2\le k \le n$, is given by,
\begin{equation}
\label{general}
s_k(y_{n-k})=\left(\sum_{j=0}^m c^{(m,\,n)}_{j,\ k}\,w^j\,\beta_{n-k}^{m-j}\right)\exp\{-(k-1)\beta_{n-k} h\}\prod_{i=n-k}^{n-2}(1-\lambda_i h).
\end{equation}
\medskip
\noindent
The proof is in the Appendix.

Now, we have a general expression for $s_{k}(y_{n-k})$ for any $2\le k \le n$, and we still need to derive a specific expression for the coefficients $\left\{c_{j,\ k}^{(m,\,n)}\right\}_{j=0}^m$. As shown by Theorem 1, the coefficients $\left\{c_{j,\ k}^{(m,\,n)}\right\}_{j=0}^m$ depend on $x^{(n)}$, where $m$ is the number of jumps in $x^{(n)}$ that has been counted in the summation so far, i.e. $m = \sum_{r_i}\mathbf1(r_i \ge n-k)$. Now we need to expand the definition of $\left\{c_{j,\ k}^{(m,\,n)}\right\}_{j=0}^m$ to $m \le \sum_{r_i}\mathbf1(r_i \ge n-k)$. 

For a sample path $x^{(n)}$, with jump times $t_i^{(n)}=(r_i+2)h$ labeled in the descending order, let's define $\left\{c_{j,\ k}^{(m,\,n)}\right\}_{j=0}^m$, with $m \le \sum_{r_i}\mathbf1(r_i \ge n-k)$, as the coefficients of $s_k\left(y_{n-k},\,x^{(m,n)}\right)$ of a sample path $x^{(m,n)}$ with jump times $\left\{t_i^{(n)}\right\}_{i=1}^m$. That means, when counting backwards along the timeline, $x^{(m,n)}$ has the first $m$ jumps same as $x^{(n)}$, but has no other jumps after then. For example, $c_{0,\ k}^{(0,\,n)}=1$ is the coefficient of $s_k\left(y_{n-k},\,x^{(0,n)}\right)$, where $x^{(0,n)}$ is a sample path with no jumps, and $\{c_{0,\ k}^{(1,\,n)},\,c_{1,\ k}^{(1,\,n)}\}$ are the coefficients of $s_k\left(y_{n-k},\,x^{(1,n)}\right)$, where $x^{(1,n)}$ is a sample path with only one jump at $t_1^{(n)}$.

In the proof of Theorem 1, we have derived a recursive equation that expresses $\left\{c_{j, \ k+1}^{(m,\,n)}\right\}_{j=0}^m$ in terms of $\left\{c_{j,\ k}^{(m,\,n)}\right\}_{j=0}^m$ when no jump happens at $x_{n-k+1}$, given by Equation~(\ref{r1}),
and a recursive equation for $\left\{c_{j,\ k+1}^{(m+1,\,n)}\right\}_{j=0}^{m+1}$ in terms of $\left\{c_{j,\ k}^{(m,\,n)}\right\}_{j=0}^m$ when a jump happens at $x_{n-k+1}$, given by Equation~(\ref{r2}).

Based on the above two recursive equations, we will be able to express $\left\{c_{j,\ k}^{(m,\,n)}\right\}_{j=0}^{m}$ in terms of $\left\{c_{j,\ k}^{(m-1,\,n)}\right\}_{j=0}^{m-1}$ for any $m>0$ and $n-r_m \le k \le n$.
This will be the key recursive relationship we use for our final result, because it gives us the way of getting $\left\{c_{j,\ n}^{(m,\,n)}\right\}_{j=0}^m$ from $c^{(0,\,n)}_{0,\ n}$ recursively for any $m>0$ and thus getting $s_n(y_0)$ recursively with any number of jumps.
This recursive equation will be given in the following theorem and proved by induction.\par
\bigskip
\noindent
\textbf{Theorem 2}. The coefficients defined in Theorem 1 have the recursive relationship from $m-1$ to $m$, given by
\begin{equation}
\label{r3}
c_{j,\ k}^{(m,\,n)}=
\begin{cases}
c_{0,\ k}^{(m-1,\,n)} &\text{for}\ j=0\\
\sum_{i=0}^{j-1} c_{i,\ k}^{(m-1,\,n)}\binom{m-i-1}{j-i-1}\sum_{i=n-k}^{r_m}\alpha_{i} h + c_{j,\ k}^{(m-1,\,n)} &\text{for}\ j=1, 2,\ldots m-1\\
\sum_{i=0}^{m-1} c_{i,\ k}^{(m-1,\,n)}\sum_{i=n-k}^{r_m}\alpha_{i} h &\text{for}\ j=m,
\end{cases}
\end{equation}
where $\alpha_{i} := e^{-(n-i-1)w h}\ \gamma_{i}$, for any $m>0$ and $n-r_m \le k \le n$.\par
\medskip
\noindent
The proof is in the Appendix.

\bigskip
Based on Section~\ref{nojump}, when there are no jumps in $x^{(n)}$, the coefficient is given by $c_{0,\ k}^{(0,\,n)} = 1$, for $2\le k\le n$. By Theorem 2, when there is only one jump in $x^{(n)}$, the coefficients are given by 
$$c_{0,\ k}^{(1,\,n)} = c_{0,\ k}^{(0,\,n)} = 1\ \ \ \ \text{and}\ \ \ \ c_{1,\ k}^{(1,\,n)} = c_{0,\ k}^{(0,\,n)}\sum_{i=n-k}^{r_1}\alpha_i h = \sum_{i=n-k}^{r_1}\alpha_i h,\ \ \ \ \text{for}\ n-r_1\le k\le n.$$
When there are two jumps in $x^{(n)}$, the coefficients are given by 
\begin{align*}
& c_{0,\ k}^{(2,\,n)} = c_{0,\ k}^{(1,\,n)} = 1 \\
& c^{(2,\,n)}_{1,\ k} = c_{0,\ k}^{(1,\,n)}
\sum_{i=n-k}^{r_2}\alpha_{i}h+c_{1,\ k}^{(1,\,n)} = \sum_{i=n-k}^{r_2}\alpha_{i}h + \sum_{i=n-k}^{r_1}\alpha_{i}h \\
& c^{(2,\,n)}_{2,\ k} = (c_{0,\ k}^{(1,\,n)} + c_{1,\ k}^{(1,\,n)})\sum_{i=n-k}^{r_2}\alpha_i h =\left( \sum_{i=n-k}^{r_1}\alpha_i h +1\right)\sum_{i=n-k}^{r_2}\alpha_i h,
\end{align*}
for $n-r_2\le k\le n$.

Therefore, based on Theorem 1, we have derived a general expression of $\text{E}_{Y^{(n)}}\left[p_{X|Y}\left(x^{(n)}\mid Y^{(n)}\right)\right]$, for the case of $m$ jumps, given by
\begin{align*}
&\text{E}_{Y^{(n)}}\left[p_{X|Y}\left(x^{(n)}\mid Y^{(n)}\right)\right] = p(x_1\mid x_0,\ y_0)s_n(y_0)\\
=&\left(\sum_{j=0}^m c^{(m,\,n)}_{j,\ n}\ w^j\beta_0^{m-j}\right)\exp(-n\beta_0 h)\prod_{i=0}^{n-2}(1-\lambda_i h),
\end{align*}
where the coefficients $\left\{c_{j,\ n}^{(m,\,n)}\right\}_{j=0}^m$ can be derived recursively from $c_{0,\ n}^{(0,\,n)}=1$ by using Theorem 2 at $k=n$.

\section{Convergence Results}\par
\label{convergence}

In this section we will prove the convergence of the expectation $$\text{E}_{Y^{(n)}}\left[p_{X|Y}\left(x^{(n)}\mid Y^{(n)}\right)\right] \rightarrow \text{E}_Y\left[p_{X|Y}(x \mid Y)\right],$$ 
by showing the convergence of $Y^{(n)}\rightarrow Y$ in distribution, and
the convergence of $x^{(n)}\rightarrow x$ in the Skorokhod topology.

\subsection{Weak Convergence}
\label{weak}
Take the non-homogeneous Poisson process $Y$ as defined in Section~\ref{model}, and its $n$-grid discretization $Y^{(n)}$ as defined in Section~\ref{discretization}.
Here we show that $Y^{(n)}\rightarrow Y$ in distribution by considering the convergence of the finite dimensional distributions; see \cite{billingsley2}.

\vspace{0.1in}
\noindent
\textbf{Lemma 1}. All the finite dimensional distributions of $Y^{(n)}$ converges weakly to the corresponding finite dimensional distributions of $Y$.

\vspace{0.1in}
\noindent
\textbf{Proof}.
For a given time point $t\in[0, T]$, let $S_t$ and $S^{(n)}_t$ denote the waiting time until the next jump for $Y$ and $Y^{(n)}$, i.e.
$$S_t = \text{inf}\{s\in[0, T-t]: Y(t+s) = Y(t) + 1\}$$
and 
$$S^{(n)}_t = \text{inf}\{s\in[0, T-t]: Y^{(n)}(t+s) = Y^{(n)}(t) + 1\}.$$
It's sufficient to show that, for any $t\in[0, T]$, $S^{(n)}_t$ converges in distribution to $S_t$, 
since both $Y$ and $Y^{(n)}$ are Markov processes.

The cumulative density functions of $S_t$ and $S^{(n)}_t$ are given by, $$F_t(s) = \text{P}(S_t \le s)\ \ \  \text{and}\ \ \ F^{(n)}_t(s)= \text{P}\left(S^{(n)}_t \le s\right),$$ 
so we want to show that $F^{(n)}_t(s)\rightarrow F_t(s)$ for any $t\in[0, T]$ and $s\in[0, T-t]$.

For $Y$, the cumulative density function of waiting time is given by, 
$$F_t(s) = \text{P}(S_t \le s) = 1 - \exp\left\{ - \int_{t}^{t+s} \gamma(u)\  du\right\}.$$
For $Y^{(n)}$, we can consider the probability that no jumps happened in $[t, t+s]$,
$$F^{(n)}_t(s)=1 - \text{P}\left(S^{(n)}_t > s \right) = 1 - \prod_{u\in \mathcal T^{(n)}\cap[t, t+s]} \left[1 - \gamma(u)\frac1n\right].$$
As $n\rightarrow\infty$, by product integral \citep{dollard2011}, the limit is given by
$$\lim_{n\rightarrow\infty}F^{(n)}_t(s) = 1 - \exp\left\{ - \int_{t}^{t+s} \gamma(u) du\right\} = F_t(s).$$
Therefore, at any given time point $t\in[0, T]$, the waiting time until the next jump of $Y^{(n)}$ converges in distribution to that of $Y$, i.e. $S^{(n)}_t \rightarrow S_t$ in distribution, which completes the proof.
\hfill$\square$

\vspace{0.2in}
\noindent
From \cite{billingsley2} Theorem 12.6, the convergence of finite dimensional distribution in $D_c$ implies weak convergence, so we have $Y^{(n)}\stackrel d\rightarrow Y$.

Now, fix $x$ and consider the conditional probability function $p_{X|Y}(x\mid y)$, given by Equation~(\ref{pxgiveny}) as a function of $y$ that maps the count path $y$ from $D_c$ to a real number. With any fixed count path $x\in D_c$, this function is 
\begin{description}
\item (i) bounded; since $x$ can only have finitely many jumps, i.e. $\{t_i\}$ is a finite set, and the intensity function $\beta(t,\ Y(t))$ is bounded;
\item (ii) continuous almost everywhere;
since the discontinuities happen only when $y$ has a jump at $t\in \{t_i\}$.

\end{description}

\noindent
Now using the continuous mapping theorem (see Theorem 2.7 in \cite{billingsley2}), with $Y^{(n)}\stackrel d\rightarrow Y$, for the bounded and continuous a.e. function $p_{X|Y}(x\mid y)$,
we have the convergence of expectations, $\text{E}_{Y^{(n)}}\left[p_{X|Y}\left(x\mid Y^{(n)}\right)\right] \rightarrow \text{E}_Y\left[p_{X|Y}(x\mid Y)\right]$.

\subsection{Convergence in Skorokhod Topology}
\label{skorokhod}
Take the count path $x$ as defined in Section~\ref{model} and its $n$-grid discretization $x^{(n)}$ as defined in Section~\ref{discretization}. 

The error in jump times induced by discretization won't be larger than the size of the time grid, i.e. $\left|t_i^{(n)} - t_i\right|<T/n$, so $t_i^{(n)}\rightarrow t_i$ as $n\rightarrow\infty$. Hence,
there exists a deformation of time scale, denoted by $\tau^{(n)}:[0, T]\rightarrow [0, T]$, which is continuous, increasing and onto, such that, $\lim_{n\rightarrow \infty}x^{(n)}\left(\tau^{(n)}(t)\right) = x(t)$ for any $t\in [0, T]$.
Furthermore,
$\left|\tau^{(n)}(t) - t\right|<T/n $, so $\tau^{(n)}(t)\rightarrow t$ uniformly and $x^{(n)}(t) \rightarrow x(t)$
for all the continuity points $t$ of $x$. Note that $x\in D_c$ only has finitely many discontinuities, so it follows $x^{(n)}\rightarrow x$ in Skorokhod Topology (see page 124 in \cite{billingsley2}). 

Consider $\text{E}_{Y}\left[p_{X|Y}(x\mid Y)\right]$ as a function of $x$ that maps $D_c[0, T]$ into $\mathbb{R}$, then this function is continuous with respect to the metric that defines the Skorokhod topolgy (12.13 in \cite{billingsley2}). Therefore, convergence of $x^{(n)}\rightarrow x$ in Skorohod topology leads to the convergence of expectations given by,
$\text{E}_{Y}\left[p_{X|Y}\left(x^{(n)}\mid Y\right)\right]\rightarrow \text{E}_{Y}\left[p_{X|Y}(x\mid Y)\right].$

\subsection{Convergence of Expectations}
\label{coe}
\noindent
\textbf{Theorem 3}.
We have the convergence of expectations given by
$$\text{E}_{Y^{(n)}}\left[p_{X|Y}\left(x^{(n)}\mid Y^{(n)}\right)\right] \rightarrow \text{E}_Y\left[p_{X|Y}(x \mid Y)\right],$$ i.e., for any $\epsilon>0$, there exist an integer $N$ such that
$$\left|\ \text{E}_{Y^{(n)}}\left[p_{X|Y}\left(x^{(n)}\mid Y^{(n)}\right)\right]\ -\  \text{E}_Y\left[p_{X|Y}(x\mid Y)\right]\ \right|<\epsilon,\ \ \ \forall\ n>N.$$
\noindent
\textbf{Proof}.
By the triangle inequality,
$\left|\ \text{E}_{Y^{(n)}}\left[p_{X|Y}\left(x^{(n)}|Y^{(n)}\right)\right]\ -\  \text{E}_Y\left[p_{X|Y}(x\mid Y)\right]\ \right|$ is upper bounded by
\begin{align*}
    \left|\ \text{E}_{Y^{(n)}}\left[p_{X|Y}\left(x^{(n)}|Y^{(n)}\right)\right]\ -\  \text{E}_{Y}\left[p_{X|Y}\left(x^{(n)}|Y\right)\right]\ \right| 
  + \left|\ \text{E}_{Y}\left[p_{X|Y}\left(x^{(n)}|Y\right)\right]\ -\ \text{E}_Y\left[p_{X|Y}(x\mid Y)\right]\ \right|.   
\end{align*}
From Section~\ref{weak}, we have $\text{E}_{Y^{(n)}}\left[p_{X|Y}\left(x\mid Y^{(n)}\right)\right] \rightarrow \text{E}_Y\left[p_{X|Y}(x\mid Y)\right]$, so for any $\epsilon>0$, there exist an integer $N_1$ such that
$$\left|\ \text{E}_{Y^{(n)}}\left[p_{X|Y}\left(x^{(n)}\mid Y^{(n)}\right)\right]\ -\  \text{E}_Y\left[p_{X|Y}\left(x^{(n)}\mid Y\right)\right]\ \right|<\epsilon/2,\ \ \ \forall\ n>N_1.$$
From Section~\ref{skorokhod}, we have $\text{E}_{Y}\left[p_{X|Y}\left(x^{(n)}|Y\right)\right]\rightarrow \text{E}_{Y}\left[p_{X|Y}(x\mid Y)\right]$, so for any $\epsilon>0$, there exist an $N_2$ such that
$$\left|\ \text{E}_{Y}\left[p_{X|Y}\left(x^{(n)}\mid Y\right)\right]\ -\  \text{E}_Y\left[p_{X|Y}(x\mid Y)\right]\ \right|<\epsilon/2,\ \ \ \forall\ n>N_2.$$
Therefore, by putting together Section \ref{weak} and \ref{skorokhod}, we have
$$\left|\ \text{E}_{Y^{(n)}}\left[p_{X|Y}\left(x^{(n)}\mid Y^{(n)}\right)\right]\ -\  \text{E}_Y\left[p_{X|Y}(x\mid Y)\right]\ \right|<\epsilon,\ \ \ \forall\ n>\max\left(N_1,N_2\right),$$
and thus
$\text{E}_{Y^{(n)}}\left[p_{X|Y}\left(x^{(n)}\mid Y^{(n)}\right)\right] \rightarrow \text{E}_Y\left[p_{X|Y}(x \mid Y)\right]$ as $n\rightarrow\infty$.
\hfill$\square$

\subsection{Marginal Likelihood}
Now, we can take the general expression of $\text{E}_{Y^{(n)}}\left[p_{X|Y}\left(x^{(n)}\mid Y^{(n)}\right)\right]$ we derived in Section~\ref{derivation}, and let $n\rightarrow\infty$ to recover the desired expectation $\text{E}_Y\left[p_{X|Y}(x \mid Y)\right]$, and thus get the marginal likelihood of the Cox process $X$ described in Section~\ref{model}.\par
\bigskip
\noindent
\textbf{Theorem 4}. Take the Cox process $\{X(t)\}_{0\le t\le T}$ and its sample path $x$ as defined in Section~\ref{model}. Suppose the sample path $x$ is characterized by the jump times $\{t_i\}_{i=1}^M$ sorted in the descending order, i.e. with $t_i>t_j$ for $i<j$. Then the marginal likelihood of $X$ is given by
$$p_X(x)=\left(\sum_{j=0}^M c_j^{(M)}w^j\beta_0^{M-j}\right) \exp\left\{-\beta_0T - \int_0^T \lambda(t)  dt \right\},$$
where,
$\lambda(t):= \left( 1 - e^{-w(T-t)}\right)\ \gamma(t),$
and the coefficients $\left\{c_j^{(M)}\right\}_{j=0}^M$ are given by the following recursive equation, for $m=1,2,\ldots,M,$
\begin{equation}
\label{r4}
\begin{split}
&c_{0}^{(0)}=1\\
&c_{j}^{(m)}=
\begin{cases}
c_{0}^{(m-1)} &\text{for}\ j=0\\
\sum_{i=0}^{j-1} c_{i}^{(m-1)}\binom{m-i-1}{j-i-1}\int_{0}^{t_m}\alpha(t)dt + c_{j}^{(m-1)} &\text{for}\ j=1, 2,\ldots m-1\\
\sum_{i=0}^{m-1} c_{i}^{(m-1)}\int_{0}^{t_m}\alpha(t)dt &\text{for}\ j=m,
\end{cases}
\end{split}
\end{equation}
with
$\alpha(t)=e^{-w\left(T-t\right)}\ \gamma(t).$\par
\medskip
\noindent
The proof is in the Appendix.

\section{Illustrations}
\label{illustration}

First, in Section~\ref{alg}, we detail the algorithms for simulating a Cox process and for calculating the marginal likelihood. Then, in Section~\ref{simu}, we present illustrations with simulated data. All posterior sampling is done using a Metropolis-Hastings algorithm which makes proposals for the parameters defining the intensity function $\gamma(t)$. The proposals are independent normal distributions with variances tuned to obtain a certain level of mixing for the chain.

We always take the intensity to be a polynomial function and so the parameters are the coefficients of the polynomial. When we simulated data it was assumed that $w$ is known. On the other hand, when we look at a real data set, see Section~\ref{real}, and transform the data to an increasing intensity function, we need to specify $w$
in a particular way and detail this in Section~\ref{real}.

The code for all illustrations in this section can be found in the web page\par
\noindent
(https://github.com/ShuyingWang/marginal-cox-process).

\begin{algorithm}
\caption{Simulate Cox process }\label{alg:simulate}
\KwIn{$T,\ \gamma(t),\ \beta_0,\ w$}
\KwOut{$\{t_i\}$}
\textbf{Define a function:} $\Gamma(t) \gets \int_0^t\gamma(t)dt$\;
Draw $u\sim$ Exponential$(1)$\;
$t^{(y)} \gets \Gamma^{-1}(u)$\;
$t^{(x)} \gets 0$\;
$y \gets 0$\;
\While{$t^{(x)} < T$}{
  Draw $r\sim$ Exponential$(\beta_0+wy)$\;
  \If{$t^{(x)}+r<\min(t^{(y)},\ T)$}{
    $t^{(x)} \gets t^{(x)}+r$\;
    Add $t^{(x)}$ into $\{t_i\}$
  }
  \ElseIf{$t^{(y)}<T$}{
    Draw $r\sim$ Exponential$(1)$\;
    $u\gets u+r$\;
    $t^{(y)} \gets \Gamma^{-1}(u)$\;
    $y\gets y+1$\;
    $t^{(x)}\gets t^{(y)}$
  }
  \Else{$t^{(x)}\gets T$}
}
\end{algorithm}

\begin{algorithm}
\caption{Calculate marginal likelihood}\label{alg:likelihood}
\KwIn{$T,\ \gamma(t),\ \beta_0,\ w,\ \{t_i\}$}
\KwOut{Marginal likelihood}
Sort $\{t_i\}$ in descending order\;
$M \gets$ count of $\{t_i\}$\;
$c_0^{(0)} \gets 1$\;
$c_0^{(1)} \gets 0$\;
\For{$m=1:M$}{
  $c_0^{(m)} \gets 1$\;
  \For{$j=1:m$}{
    $c_j^{(m)} \gets \sum_{i=0}^{j-1} c_{i}^{(m-1)}\binom{m-i-1}{j-i-1}\int_0^{t_m}e^{-w\left(T-t\right)}\gamma(t)\,dt + c_{j}^{(m-1)}$\;
  }
  $c_{m+1}^{(m)} \gets 0$\;
}
Marginal likelihood $\gets\left(\sum_{j=0}^M c_j^{(M)}w^j\beta_0^{M-j}\right) \exp\left\{-\beta_0T - \int_0^T \left( 1 - e^{-w(T-t)}\right)\gamma(t)\,dt \right\}$\;
\end{algorithm}

\subsection{Algorithms}
\label{alg}

To simulate the Cox process defined in Section~\ref{model}, we use the standard approach of simulating non-homogeneous Poisson processes; see \cite{kingman1992}. 
First, we need to simulate the latent non-homogeneous Poisson process $\{Y(t)\}_{0\le t\le T}$ with the intensity function $\gamma(t)$, by converting from a standard Poisson process. Then, given a sample path of $Y$, we simulate $X$ as a non-homogeneous Poisson process with the intensity $(\beta_0 + wY(t))$; see Algorithm~\ref{alg:simulate}.
The marginal likelihood computation using the recursive expression as in Theorem 4, is written out in  Algorithm~\ref{alg:likelihood}.

\subsection{Simulated Data}
\label{simu}

First, we do some illustrations with simulated data and the intensity $\gamma(t)$ being polynomial functions. In particular, we assume the intensity function ranges over 4 types, namely constant, linear, quadratic and cubic. In each case the data were generated using Algorithm 1. 
For all cases, independent normal distributions with large variances are used as weak priors.
Also, in each case, a Metropolis sampler was used to obtain samples from the posterior distribution of the coefficients of the polynomial function. The variances of the independent normal proposals are tuned to obtain a 0.2 to 0.3 acceptance rate when an entire new function is proposed. That is either all the parameters move or the function remains as at the current state.

\begin{figure}[h!]
\begin{subfigure}{0.5\textwidth}
\includegraphics[width=8cm, height=8cm]{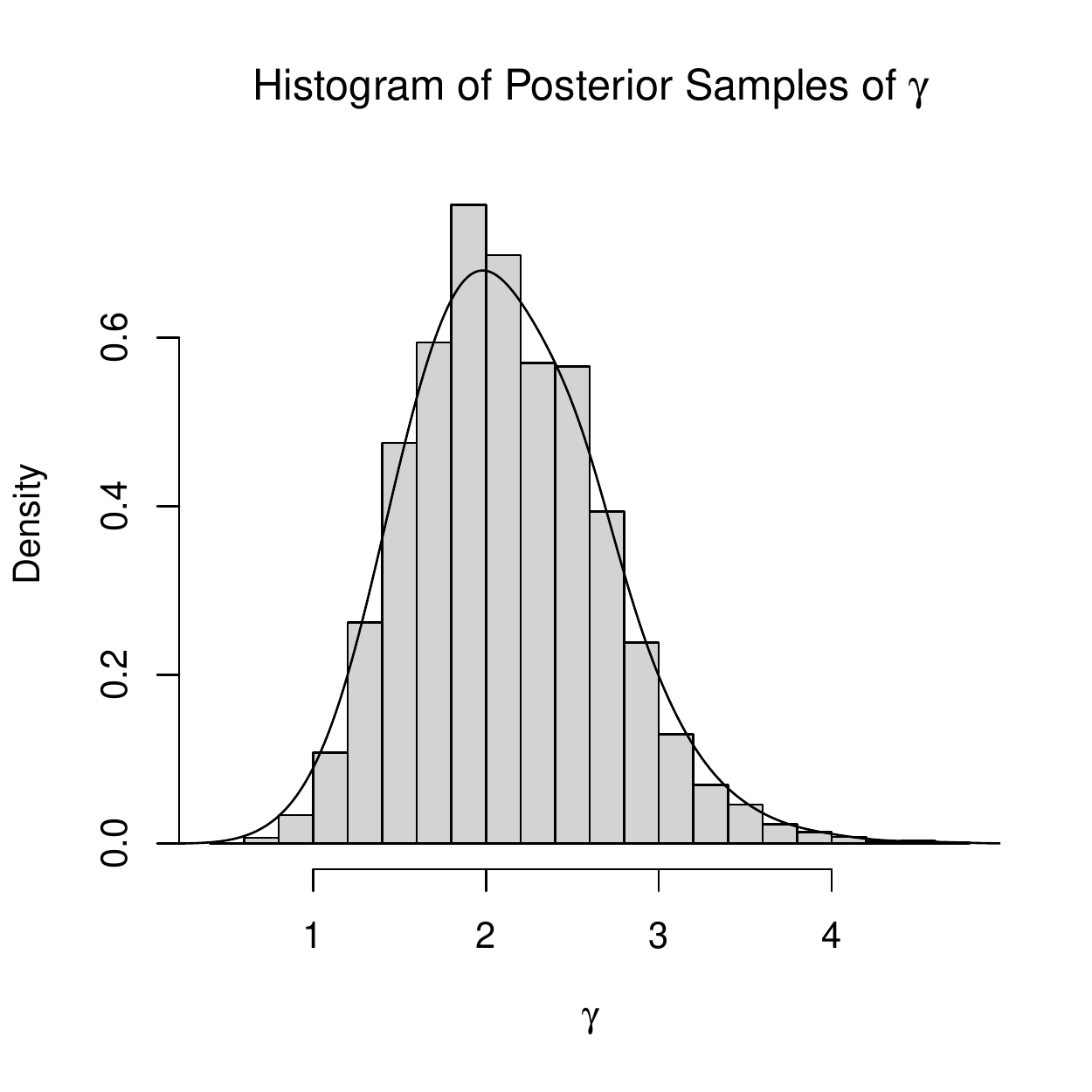} 
\caption{Constant $\gamma$}
\label{fig1.1}
\end{subfigure}
\begin{subfigure}{0.5\textwidth}
\includegraphics[width=8cm, height=8cm]{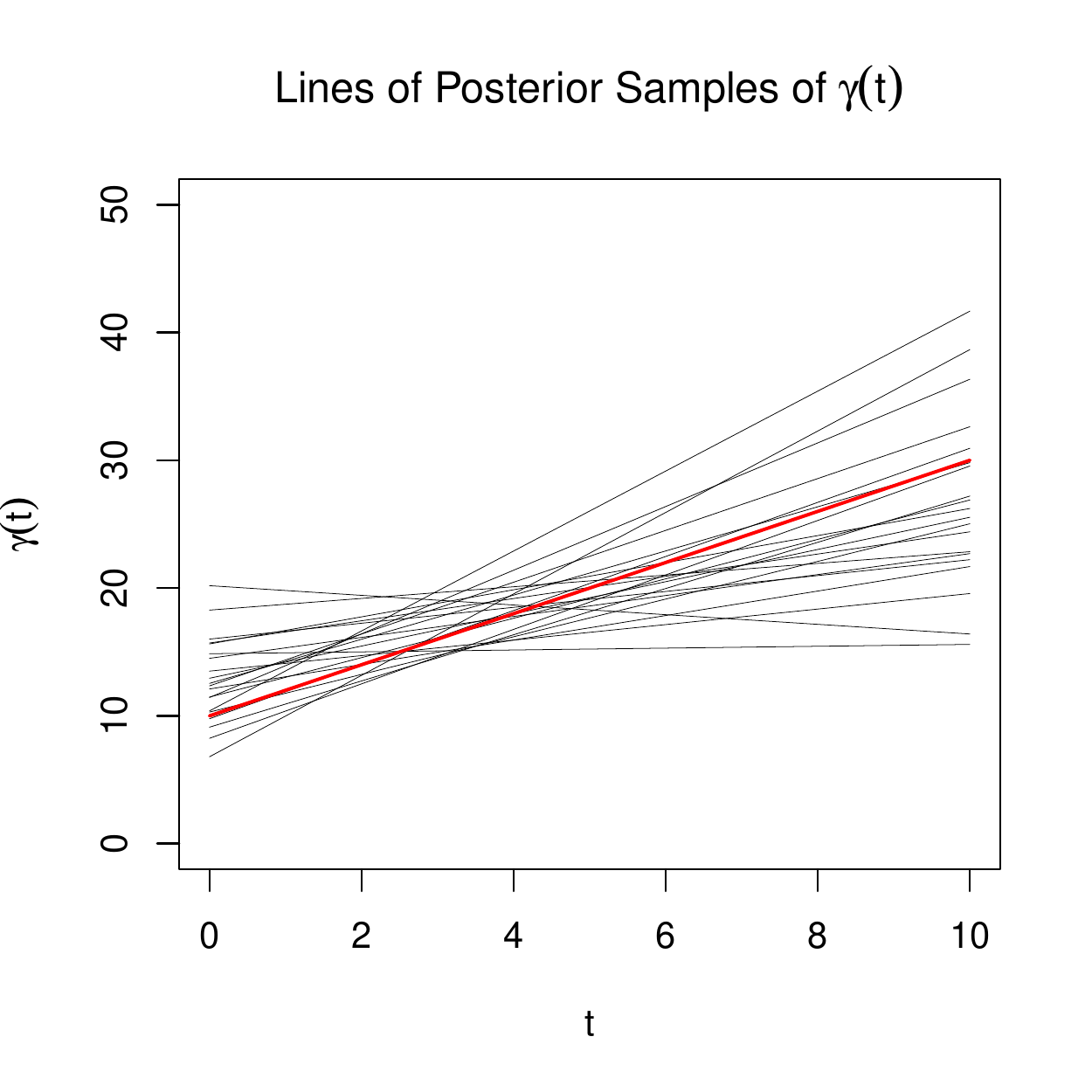}
\caption{Linear $\gamma(t)$}
\label{fig1.2}
\end{subfigure}
\begin{subfigure}{0.5\textwidth}
\includegraphics[width=8cm, height=8cm]{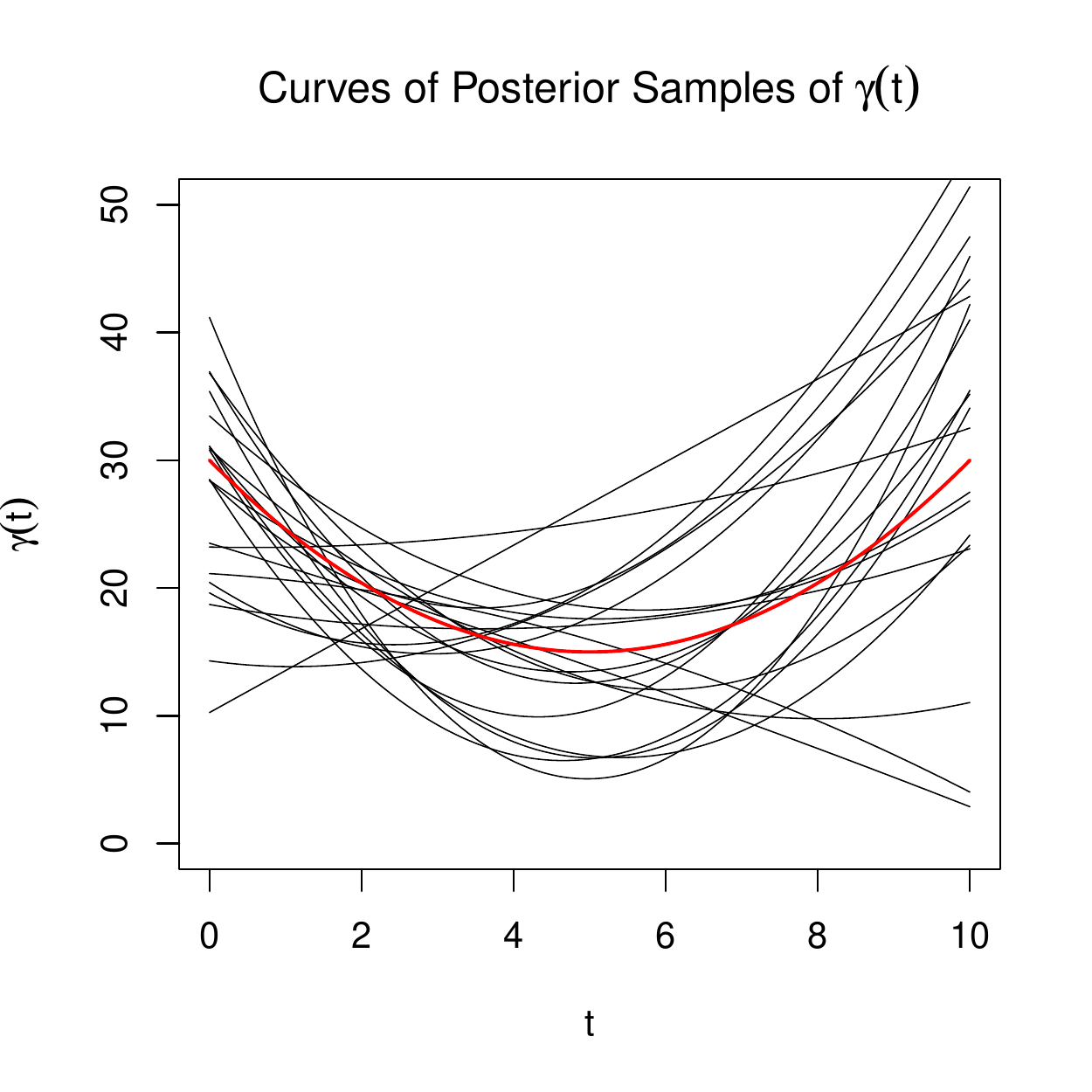}
\caption{Quadratic $\gamma(t)$}
\label{fig1.3}
\end{subfigure}
\begin{subfigure}{0.5\textwidth}
\includegraphics[width=8cm, height=8cm]{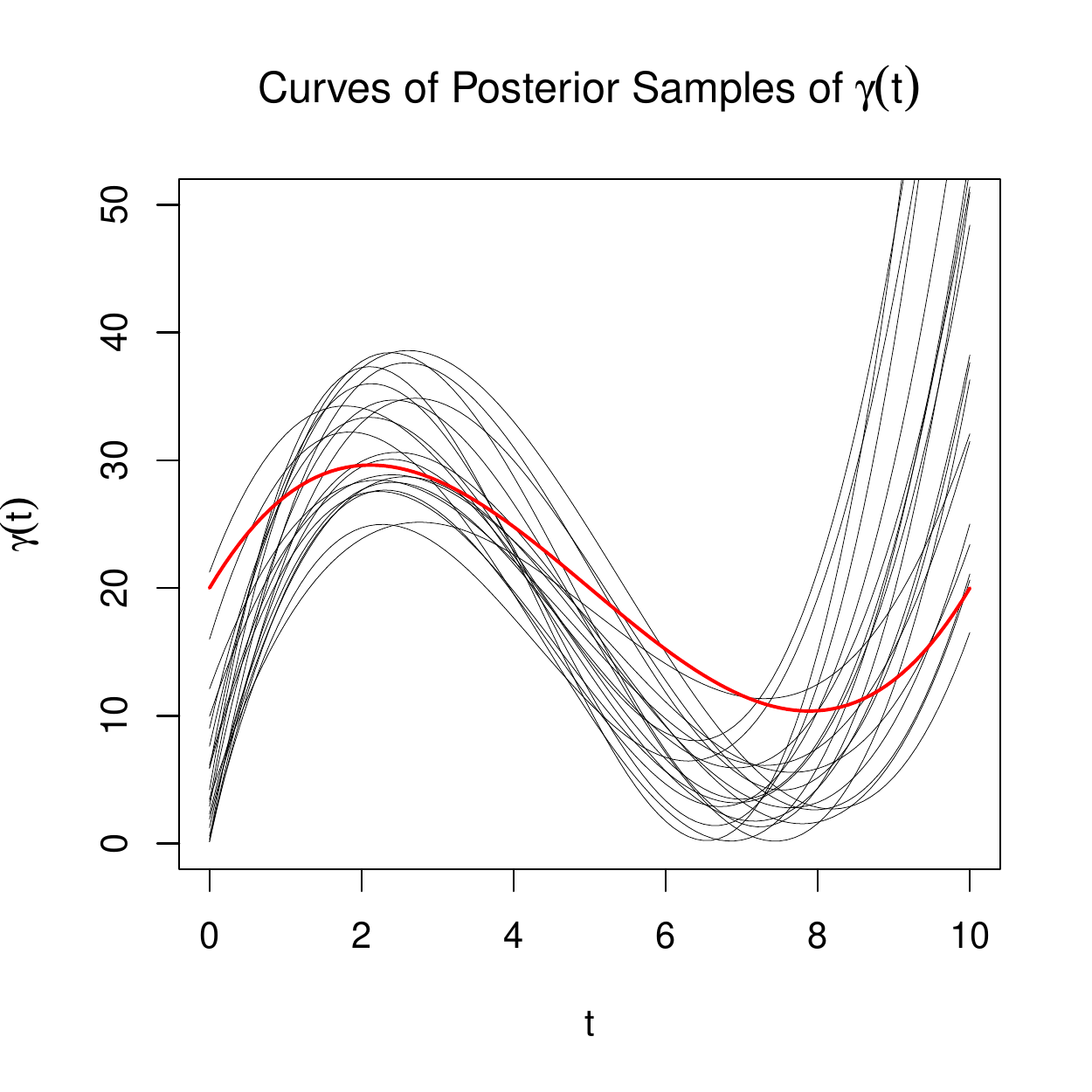}
\caption{Cubic $\gamma(t)$}
\label{fig10.2}
\end{subfigure}
\caption{Simulated data example: The four figures are samples from the posterior for the constant, linear, quadratic and cubic polynomials. For the first case the posterior samples are the constant value and for the last three we present random polynomial curves from the posterior with the true curve of $\gamma(t)$ in red.}
\label{fig1}
\end{figure}

Sample functions alongside the true function in each case are presented in Fig.~\ref{fig1}. In Fig.~\ref{fig1.1} the function is represented by a constant and the posterior samples are collected in a histogram, overlapped with an empirical density curve. The true value of $\gamma$ is 2, with the posterior mean being 2.12, and the posterior mode being 1.98.
In all others; i.e. Fig.~\ref{fig1.2}, Fig.~\ref{fig1.3}, and Fig.~\ref{fig10.2}, a number of posterior functions are plotted alongside the true function of $\gamma(t)$ which is in red.

\subsection{Real Data}
\label{real}

For the real data illustration, we use the AEGISS dataset, which is originally from \cite{diggle2005data}, and documented in \citet{taylor2013}. This is a spatio-temporal disease surveillance dataset that records the time and location of disease events. 
In this paper, we only focus on time series data, so we take a rectangle area from the space of the dataset, and take the time points of events happened in this area as a count process. In order to fit our Cox process model, we transform the original count process into an adapted count process with non-decreasing intensity. The way of adapting the data is as follows.
Suppose the original count process has $M^\star$ jumps in the time interval $[0, T]$. Let $\left\{t_i^\star\right\}_{i=1}^{M^\star}$ denote the jump times and $x^\star(t)=\sum_{i=1}^{M^\star}\mathbf1(t_i\le t)$ be the sample path of the original count process.
First, we take the integral of $x^\star(t)$ and scale it with a constant $w$ to get an integral function given by $\Tilde{x}(t)=w\int_0^t x^\star(s)ds$. Second, we take the time points when $\Tilde{x}(t)$ first reaches each integer value, denoted as $\left\{\Tilde{t_i}\right\}_{i=1}^M$, so $\Tilde{t_i} = \inf\{t: \Tilde{x}(t)=i\}$. 
Third, we convert $\Tilde{x}(t)$ into a count path $x(t)$ with $M$ jumps, such that $x(t)$ coincides with $\Tilde{x}(t)$ at each $\Tilde{t_i}$, and the area under $x(t)$ equals the area under $\Tilde{x}(t)$ up to each $\Tilde{t_i}$, i.e. $x\left(\Tilde{t_i}\right)=\Tilde{x}\left(\Tilde{t_i}\right)$ and $\int_0^{\Tilde{t_i}}x\left(t\right)dt=\int_0^{\Tilde{t_i}}\Tilde{x}\left(t\right)dt$ for $i=1,\ldots,M$. Additionally, we can adjust the value of $w$ to obtain $M=M^\star$, so that the adapted count process $x(t)$ has the same number of jumps as the original count process $x^\star(t)$.

After adapting the time series data, we can fit the Cox process model with $\beta_0=0$ and with the same $w$ value we used for scaling the integral. We use polynomial $\gamma(t)$ functions and weak independent normal priors for the coefficients and use Metropolis samplers with independent normal proposals to obtain the posterior samples of the coefficients. Finally, for illustrative purposes, we use the posterior mean of each coefficient to get an estimate of the $\gamma(t)$ function.

\begin{figure}[h!]
\begin{subfigure}{0.33\textwidth}
\includegraphics[width=6cm, height=6cm]{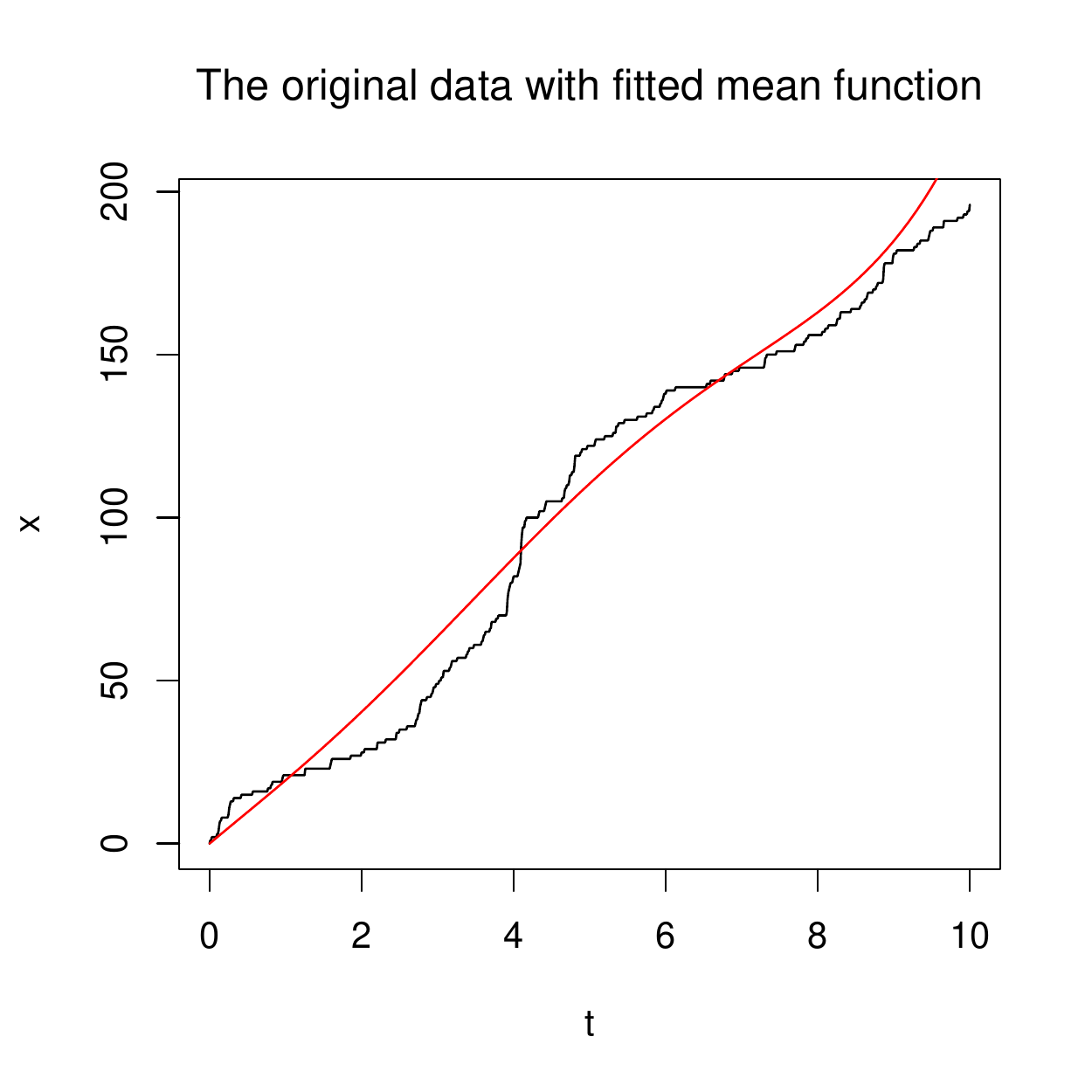} 
\caption{4th order polynomial $\gamma(t)$}
\label{fig2.1}
\end{subfigure}
\begin{subfigure}{0.33\textwidth}
\includegraphics[width=6cm, height=6cm]{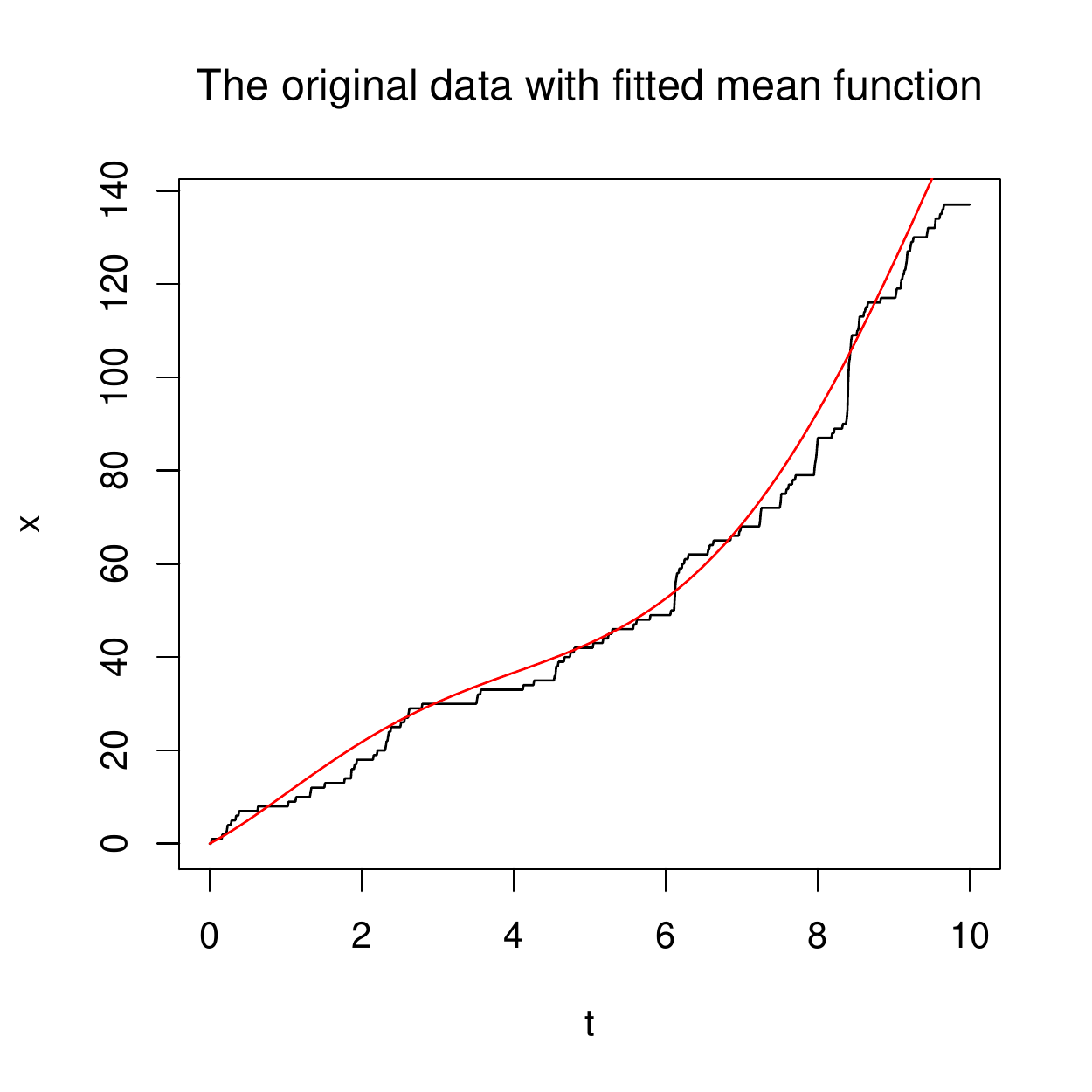}
\caption{4th order polynomial $\gamma(t)$}
\label{fig2.2}
\end{subfigure}
\begin{subfigure}{0.33\textwidth}
\includegraphics[width=6cm, height=6cm]{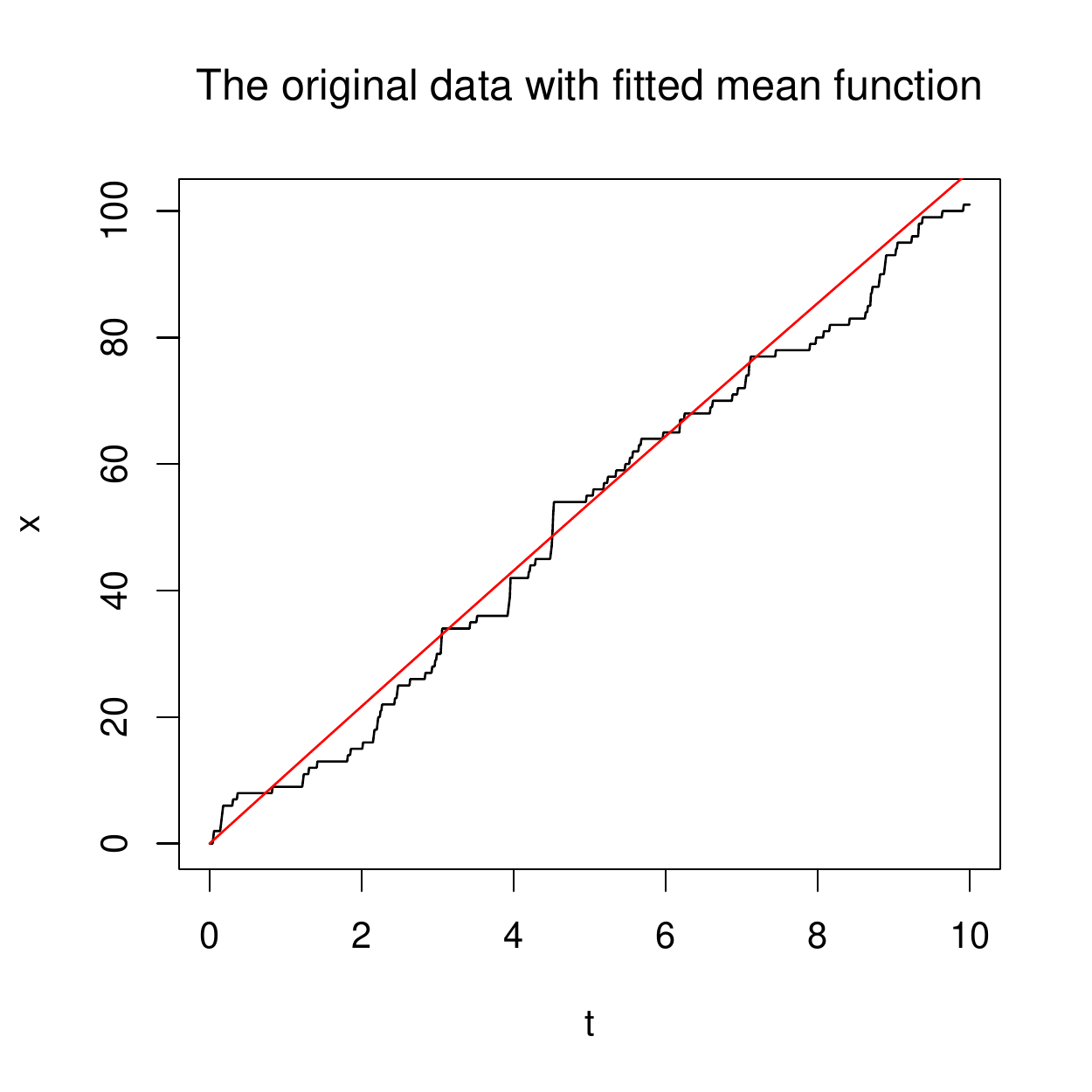}
\caption{Linear $\gamma(t)$}
\label{fig2.3}
\end{subfigure}
\caption{Real data example: The three figures are time series data of three different rectangle areas selected from the original spatio-temporal dataset. The curve in red is the integral of the fitted $\gamma(t)$ function}
\label{fig2}
\end{figure}

We repeat the above procedures for three different rectangle areas selected from the original dataset, and obtain the fitted polynomial $\gamma(t)$ as the mean intensity function for each time series; see Fig.~\ref{fig2}.

\section{Discussion}
\label{discuss}

In the paper we have constructed a non-homogeneous count process. This is done in a well motivated way by first setting up a latent intensity process for a non-homogeneous Poisson process and then integrating it out to obtain the marginal process. This, in short, is a marginalized Cox process. As a consequence, direct likelihood computations are available without approximation or the need to sample latent processes. While we have used low dimensional polynomials to demonstrate the approach in the current paper, extending to more general types of function is quite possible and simple to do; the family of functions just needs to be integrable on $[0,T]$. Future work will involve how we can extend the marginalization in a multidimensional setting.

\bibliographystyle{apalike}

\bibliography{Bib}

\newpage

\section*{Appendix}

\noindent
\textbf{Proof of Theorem 1}. To prove this theorem by induction, we first show that it's true for the base case at $k=2$, then we show that if it holds for $k$, it will also hold for $k+1$.

First, for the base case at $k=2$, we have it derived in Section~\ref{nojump}, given by
$$s_2(y_{n-2})=\exp(-\beta_{n-2} h) \ (1 - \lambda_{n-2} h),$$
which follows the general expression given by Equation~(\ref{general}), with $m=0$ and $c_{0,\ 2}^{(0,\,n)}=1$.

Next, to show the induction step, suppose there exists a $2\le k < n$, such that $s_k(y_{n-k})$ follows the general expression. Then, we need to discuss two different cases for $s_{k+1}(y_{n-k-1})$, depending on whether or not there is a jump at $x_{n-k+1}$.

When there is no jump at $x_{n-k+1}$, which means when $t=(n-k+1)h$ is not a time point of jump, i.e. $n-k+1\ne r_{m+1}+2$, we have $x_{n-k+1}=x_{n-k}$ and
\begin{align*}
s_{k+1}(y_{n-k-1}) = &\sum_{y_{n-k}} p(x_{n-k+1}\mid x_{n-k},\,y_{n-k})\,\text{P}(y_{n-k}\mid y_{n-k-1}) \,s_{k}(y_{n-k}) \\
=&\sum_{j=0}^m c^{(m,\,n)}_{j,\ k}\,w^j\left\{\beta_{n-k-1}^{m-j}(1-\gamma_{n-k-1} h)+(\beta_{n-k-1}+w)^{m-j}e^{-kwh}\gamma_{n-k-1} h\right\}\\
&\times \exp(-k\beta_{n-k-1} h)\prod_{i=n-k}^{n-2}(1-\lambda_i h)
\end{align*}
When there is a jump at $x_{n-k+1}$, which means when $t=(n-k+1)h$ is a time point of jump i.e. $n-k+1=r_{m+1}+2$, we have $x_{n-k+1}=x_{n-k}+1$ and
\begin{align*}
&s_{k+1}(y_{n-k-1})\\
=&\sum_{j=0}^m c^{(m,\,n)}_{j,\ k}\,w^j\left\{\beta_{n-k-1}^{m+1-j}(1-\gamma_{n-k-1} h)+(\beta_{n-k-1}+w)^{m+1-j}e^{-kwh}\gamma_{n-k-1} h\right\}\\
&\times \exp(-k\beta_{n-k-1} h)\prod_{i=n-k}^{n-2}(1-\lambda_i h).
\end{align*}
Here we omit the subscripts to show the following algebraic steps more clearly,
\begin{equation*}
\begin{split}
&\beta^{i}\ (1 - \gamma h)\ +\ (\beta+w)^{i}\ e^{-kw h}\ \gamma h\\
=&\beta^{i}\left\{1 - \left( 1 - e^{-kw h}\right)\gamma h  \right\}+\{(\beta+w)^{i} - \beta^{i}\}\ e^{-kw h}\ \gamma h\\
=&\left[\beta^{i}+\{(\beta+w)^{i} - \beta^{i}\}e^{-kw h}\ \gamma h\right]\left\{1 - \left( 1 - e^{-kw h}\right)\gamma h  \right\} + o(h)\\
=&\left[\beta^{i}+\{(\beta+w)^{i} - \beta^{i}\}\ e^{-kw h}\ \gamma h\right](1 - \lambda h).
\end{split}
\end{equation*}
The $h^2$ term can be considered as an $o(h)$, which will become $0$ as we set $n\rightarrow \infty$, so we can omit such $o(h)$ terms in the rest of this section.
Define $\alpha_{i} := e^{-(n-i-1)w h}\ \gamma_{i}$, and put things together, we can get
\begin{align*}
s_{k+1}(y_{n-k-1})=
\begin{cases}
&\sum_{j=0}^m c^{(m,\,n)}_{j,\ k}\ w^j\big[\beta^{m-j} + \{(\beta+w)^{m-j} - \beta^{m-j}\}\alpha_{n-k-1} h\big] \\
&\times\exp(-k\beta h)\prod_{i=n-k-1}^{n-2}(1-\lambda_i h) \ \ \ \ \ \ \ \ \ \ \ \text{for}\ x_{n-k+1} = x_{n-k}\\
& \\
&\sum_{j=0}^m c^{(m,\,n)}_{j,\ k}\ w^j\big[\beta^{m+1-j} + \{(\beta+w)^{m+1-j} - \beta^{m+1-j}\}\alpha_{n-k-1} h\big] \\
&\times\exp(-k\beta h)\prod_{i=n-k-1}^{n-2}(1-\lambda_i h) \ \ \ \ \ \ \ \ \ \ \ \text{for}\ x_{n-k+1} = x_{n-k}+1,
\end{cases}
\end{align*}
where the subscripts of $\beta$, given by $n-k-1$, are omitted.
We can see that, we do have the ``$e^{-k\beta h}\prod(1-\lambda_i h)$" part that follows the general expression given by Equation~(\ref{general}), so now we only focus on the polynomial part of $s_{k+1}(y_{n-k-1})$.

When there is no jump at $x_{n-k+1}$, the polynomial part of $s_{k+1}(y_{n-k-1})$ is given by
\begin{align*}
&\sum_{i=0}^{m} c^{(m,\,n)}_{i,\ k}\ w^i\left[\beta^{m-i} + \{(\beta+w)^{m-i} - \beta^{m-i}\}\alpha h\right] \\
= &\sum_{i=0}^{m} c^{(m,\,n)}_{i,\ k}\ w^i\left[\beta^{m-i} + \left\{\sum_{l=1}^{m-i}\binom{m-i}{l}w^l\beta^{m-i-l}\right\}\alpha h\right]\\
= &\sum_{j=0}^m  c_{j,\ k+1}^{(m,\,n)} w^j\beta^{m-j}.
\end{align*}
By matching the coefficients, we can get
\begin{equation}
\label{r1}
c_{j,\ k+1}^{(m,\,n)} =
\begin{cases}
c_{0,\ k}^{(m,\,n)} &\text{for}\ j=0 \\
\sum_{i=0}^{j-1} c_{i,\ k}^{(m,\,n)}\binom{m-i}{j-i}\alpha_{n-k-1} h + c_{j,\ k}^{(m,\,n)} &\text{for}\ j=1,2,\ldots m,
\end{cases}
\end{equation}
for $x_{n-k+1} = x_{n-k}$.

When there is a jump at $x_{n-k+1}$, the polynomial part of $s_{k+1}(y_{n-k-1})$ is given by
\begin{align*}
&\sum_{i=0}^{m} c^{(m,\,n)}_{i,\ k}\ w^i\left[\beta^{m+1-i} + \{(\beta+w)^{m+1-i} - \beta^{m+1-i}\}\alpha h\right] \\
= &\sum_{j=0}^{m+1}  c_{j,\ k+1}^{(m+1,\,n)} \, w^j\,\beta^{m+1-j}.
\end{align*}
By matching the coefficients, we can get
\begin{equation}
\label{r2}
c_{j,\ k+1}^{(m+1,\,n)} =
\begin{cases}
c_{0,\ k}^{(m,\,n)} &\text{for}\ j=0 \\
\sum_{i=0}^{j-1} c_{i,\ k}^{(m,\,n)}\binom{m+1-i}{j-i}\alpha_{n-k-1} h + c_{j,\ k}^{(m,\,n)} &\text{for}\ j=1,2,\ldots m\\
\sum_{i=0}^{m} c_{i,\ k}^{(m,\,n)}\ \alpha_{n-k-1} h &\text{for}\ j=m+1,
\end{cases}
\end{equation}
for $x_{n-k+1} = x_{n-k}+1$.
Therefore, for both cases, $s_{k+1}(y_{n-k-1})$ follows the general expression given by Theorem 1, which completes the proof.
\hfill$\square$\par
\vspace{0.2in}
\noindent
\textbf{Proof of Theorem 2}. To prove this theorem by induction, we first show that it's true for the base case at $k=n-r_m$, then we show that if it holds for $k$, it will also hold for $k+1$. 

In this proof, we will omit the case of $j=0$, which is trivial to show, and define $c_{m,\ k}^{(m-1,\,n)}:=0$ to combine the case of $j=1,\ldots,m-1$ and $j=m$.\par
\medskip
First, we need to show the base case at $k=n-r_m$. Here we have a jump at $x_{r_m + 2}$, which is the $m$th jump that has been counted in the summation, so we can use Equation~(\ref{r2}) to get,
\begin{equation*}
\begin{split}
c_{j,\ n-r_m}^{(m,\,n)} &=\sum_{i=0}^{j-1} c_{i,\ n-r_m-1}^{(m-1,\,n)}\binom{m-i}{j-i}\alpha_{r_m} h + c_{j,\ n-r_m-1}^{(m-1,\,n)}\\
&=\sum_{i=0}^{j-1} c_{i,\ n-r_m-1}^{(m-1,\,n)}\left\{\binom{m-i-1}{j-i-1}+\binom{m-i-1}{j-i}\right\}\alpha_{r_m} h + c_{j,\ n-r_m-1}^{(m-1,\,n)},
\end{split}
\end{equation*}
for $j=1,\ldots m$.

Then, by Equation~(\ref{r1}), we can substitute the $c_{j,\ n-r_m-1}^{(m-1,\,n)}$ in the above equation by
$$c_{j,\ n-r_m-1}^{(m-1,\,n)}=c_{j,\ n-r_m}^{(m-1,\,n)} -\sum_{i=0}^{j-1} c_{i,\ n-r_m-1}^{(m-1,\,n)}\binom{m-1-i}{j-i}\alpha_{r_m} h.$$
The $\left\{c_{i,\ n-r_m-1}^{(m-1,\,n)}\right\}_{i=0}^{j-1}$ can be directly replaced with $\left\{c_{i,\ n-r_m}^{(m-1,\,n)}\right\}_{i=0}^{j-1}$ by adding an $o(h)$ term. Therefore, we can get
\begin{align*}
&c_{j,\ n-r_m}^{(m,\,n)}=\sum_{i=0}^{j-1} c_{i,\ n-r_m}^{(m-1,\,n)}\binom{m-i-1}{j-i-1}\alpha_{r_m} h + c_{j,\ n-r_m}^{(m-1,\,n)},\ \ \ \text{for}\ j=1,2,\ldots m,
\end{align*}
which completes the proof of the base case at $k=n-r_m$.\par
\medskip
Next, we need to show the induction step, that when Equation~(\ref{r3}) holds for $\left\{c_{j,\ k}^{(m,\,n)}\right\}_{j=0}^{m}$, it will also hold for $\left\{c_{j,\ k+1}^{(m,\,n)}\right\}_{j=0}^{m}$.

Now, suppose we have a $n-r_m \le k \le n$ such that the Equation~(\ref{r3}) holds. By Equation~(\ref{r1}), we can get
$$c_{j,\ k+1}^{(m,\,n)}=\sum_{i=0}^{j-1} c_{i,\ k}^{(m,\,n)}\binom{m-i}{j-i}\alpha_{n-k-1} h + c_{j,\ k}^{(m,\,n)},\ \ \ \text{for}\ j=1,2,\ldots,m.$$
Then we can substitute the $\left\{c_{i,\ k}^{(m,\,n)}\right\}_{i=0}^j$ in the above equation by using Equation~(\ref{r3}), since we have assumed that it holds for $k$,
\begin{align*}
c_{j,\ k+1}^{(m,\,n)}=&\sum_{i=1}^{j-1} \left\{\sum_{l=0}^{i-1} c_{l,\ k}^{(m-1,\,n)}\binom{m-l-1}{i-l-1}\sum_{l=n-k}^{r_m}\alpha_{l} h + c_{i,\ k}^{(m-1,\,n)}\right\}\binom{m-i}{j-i}\alpha_{n-k-1} h \\
&+c_{0,\ k}^{(m-1,\,n)}\binom{m}{j}\alpha_{n-k-1} h+ \sum_{i=0}^{j-1} c_{i,\ k}^{(m-1,\,n)}\binom{m-i-1}{j-i-1}\sum^{r_m}_{i=n-k}\alpha_{i} h + c_{j,\ k}^{(m-1,\,n)}\\
=&\sum_{i=1}^{j-1} \left\{\sum_{l=0}^{i-1} c_{l,\ k}^{(m-1,\,n)}\binom{m-l-1}{i-l-1} \right\}\binom{m-i}{j-i}\alpha_{n-k-1} h \sum_{i=n-k-1}^{r_m}\alpha_{i} h + o(h)\\
& + \sum_{i=0}^{j-1} c_{i,\ k}^{(m-1,\,n)}\left\{\binom{m-i-1}{j-i-1}+\binom{m-i-1}{j-i} \right\}\alpha_{n-k-1} h\\
&+ \sum_{i=0}^{j-1} c_{i,\ k}^{(m-1,\,n)}\binom{m-i-1}{j-i-1}\sum_{i=n-k}^{r_m}\alpha_{i} h + c_{j,\ k}^{(m-1,\,n)}.
\end{align*}
Then, we use Equation~(\ref{r1}) to substitute the $c_{j,\ k}^{(m-1,\,n)}$ in the above equation by,
$$c_{j,\ k}^{(m-1,\,n)}=c_{j,\ k+1}^{(m-1,\,n)} -\sum_{i=0}^{j-1} c_{i,\ k}^{(m-1,\,n)}\binom{m-1-i}{j-i}\alpha_{n-k-1} h,$$
and we will get
\begin{align*}
c_{j,\ k+1}^{(m,\,n)}
=&\sum_{i=1}^{j-1} \left\{\sum_{l=0}^{i-1} c_{l,\ k}^{(m-1,\,n)}\binom{m-l-1}{i-l-1} \right\}\binom{m-i}{j-i}\alpha_{n-k-1} h \sum_{i=n-k}^{r_m}\alpha_{i} h\\
&+ \sum_{i=0}^{j-1} c_{i,\ k}^{(m-1,\,n)}\binom{m-i-1}{j-i-1}\sum_{i=n-k-1}^{r_m}\alpha_{i} h + c_{j,\ k+1}^{(m-1,\,n)}.
\end{align*}
Then, we use Equation~(\ref{r1}) again to substitute the $\left\{c_{i,\ k}^{(m-1,\,n)}\right\}_{i=0}^{j-1}$ in the above equation and get
\begin{align*}
c_{j,\ k+1}^{(m,\,n)}
=&\sum_{i=1}^{j-1} \left\{\sum_{l=0}^{i-1} c_{l,\ k}^{(m-1,\,n)}\binom{m-l-1}{i-l-1} \right\}\binom{m-i}{j-i}\alpha_{n-k-1} h \sum_{i=n-k-1}^{r_m}\alpha_{i} h\\
&+ \sum_{i=1}^{j-1} \left\{c_{i,\ k+1}^{(m-1,\,n)} -\sum_{l=0}^{i-1} c_{l,\ k}^{(m-1,\,n)}\binom{m-l-1}{i-l}\alpha_{n-k-1} h\right\}\binom{m-i-1}{j-i-1}\sum_{i=n-k-1}^{r_m}\alpha_{i} h \\
& + c_{0,\ k+1}^{(m-1,\,n)}\binom{m-1}{j-1}\sum_{i=n-k-1}^{r_m}\alpha_{i} h
+ c_{j,\ k+1}^{(m-1,\,n)}\\
=&\sum_{l=0}^{j-2} c_{l,\ k}^{(m-1,\,n)}\left\{\sum_{i=l+1}^{j-1} \binom{m-l-1}{i-l-1}\binom{m-i}{j-i}\right\}\alpha_{n-k-1} h \sum_{i=n-k-1}^{r_m}\alpha_{i} h \\
&- \sum_{l=0}^{j-2}c_{l,\ k}^{(m-1,\,n)} \left\{\sum_{i=l+2}^{j} \binom{m-l-1}{i-l-1}\binom{m-i}{j-i}\right\}\alpha_{n-k-1} h\sum_{i=n-k-1}^{r_m}\alpha_{i} h \\
&+\sum_{i=0}^{j-1} c_{i,\ k+1}^{(m-1,\,n)}
\binom{m-i-1}{j-i-1}\sum_{i=n-k-1}^{r_m}\alpha_{i} h + c_{j,\ k+1}^{(m-1,\,n)}\\
=&\sum_{i=0}^{j-1} c_{i,\ k+1}^{(m-1,\,n)}\binom{m-i-1}{j-i-1}\sum_{i=n-k-1}^{r_m}\alpha_{i} h + c_{j,\ k+1}^{(m-1,\,n)},
\end{align*}
which completes the proof.
\hfill$\square$

\vspace{0.2in}
\noindent
\textbf{Proof of Theorem 4}. Based on Section~\ref{mjumps} and Section~\ref{coe}, we have
\begin{align*}
p_X(x)=&\text{E}_Y\left[p_{X|Y}(x \mid Y)\right]=\lim_{n\rightarrow\infty}\text{E}_{Y^{(n)}}\left[p_{X|Y}\left(x^{(n)}\mid Y^{(n)}\right)\right] \\
=&\lim_{n\rightarrow\infty}\left(\sum_{j=0}^M c^{(M,\,n)}_{j,\ n}\ w^j\beta_0^{M-j}\right)\exp(-n\beta_0 h)\prod_{i=0}^{n-2}(1-\lambda_i h)\\
=&\left(\sum_{j=0}^M \lim_{n\rightarrow\infty}c^{(M,\,n)}_{j,\ n}\ w^j\beta_0^{M-j}\right)\exp(-\beta_0 T)\lim_{n\rightarrow\infty}\prod_{i=0}^{n-2}(1-\lambda_i h).
\end{align*}

By product integral \citep{dollard2011}, we have
\begin{align*}
\lim_{n\rightarrow\infty}\prod_{i=0}^{n-2}(1-\lambda_i h)=&\lim_{n\rightarrow\infty}\prod_{i=0}^{n-2}\left\{1-\left(1 - e^{-(n-i-1)w h}\right) \gamma_i h\right\}\\
=&\lim_{h\rightarrow0}\prod_{t=0}^{T-2h}\left\{1-\left(1 - e^{-w(T-t-h)}\right) \gamma(t)h\right\}\\
=&-\int_0^T\left( 1 - e^{-w(T-t)}\right)\gamma(t)\, dt
\end{align*}
Now, we need to show that as $n\rightarrow\infty$, the $c^{(m,\,n)}_{j,\ n}$ given in Equation~(\ref{r3}) converges to the $c^{(m)}_{j}$ given in Equation~(\ref{r4}) for any $m\ge0$  and $j=0,\ldots,m$, which can be proved by induction. 
For the base case at $m=0$, it is obvious that $\lim_{n\rightarrow\infty}c^{(0,\,n)}_{0,\ n}=c_0^{(0)}=1$. 

For the induction step, we need to show that $\lim_{n\rightarrow\infty}c^{(m-1,\,n)}_{j,\ n}=c_j^{(m-1,\,n)}$ implies $\lim_{n\rightarrow\infty}c^{(m,\,n)}_{j,\ n}=c_j^{(m)}$.
Again, here we will omit the case of $j=0$, which is trivial to show, and define $c_{m}^{(m-1)}:=0$ to combine the case of $j=1,\ldots,m-1$ and $j=m$. By Equation~(\ref{r3}), we have
\begin{align*}
\lim_{n\rightarrow\infty}c_{j,\ n}^{(m,\,n)}=&
\lim_{n\rightarrow\infty}\left\{\sum_{i=0}^{j-1} c_{i,\ n}^{(m-1,\,n)}\binom{m-i-1}{j-i-1}\sum_{i=0}^{r_m}\alpha_{i} h + c_{j,\ n}^{(m-1,\,n)}\right\}\\
=&\sum_{i=0}^{j-1} \lim_{n\rightarrow\infty} c_{i,\ n}^{(m-1,\,n)}\binom{m-i-1}{j-i-1}\left\{\lim_{n\rightarrow\infty}\sum_{i=0}^{r_m}\alpha_{i} h\right\} + \lim_{n\rightarrow\infty}c_{j,\ n}^{(m-1,\,n)}.
\end{align*}
Based on Section~\ref{skorokhod}, we have $t_i^{(n)}\rightarrow t_i$ as $n\rightarrow\infty$, so by Riemann integral, we can get
$$\lim_{n\rightarrow\infty}\sum_{i=0}^{r_m}\alpha_i h=\lim_{h\rightarrow0}\sum_{t=0}^{t_m-2h}e^{-(T-t-h)w}\,\gamma(t)\,h=\int_{0}^{t_m}e^{-(T-t)w}\,\gamma(t)\,dt.$$
Therefore, if we have
$\lim_{n\rightarrow\infty}c^{(m-1,\,n)}_{j,\ n}=c_j^{(m-1)}$ for $j=0,\ldots,m-1$, then we will also have $\lim_{n\rightarrow\infty}c^{(m,\,n)}_{j,\ n}=c_j^{(m)}$ for $j=0,\ldots,m$, which completes the proof.
\hfill$\square$

\end{document}